\documentclass[twocolumn,amssymb,prl,aps]{revtex4}
\input psfig
\newenvironment{myitemize} {
  \begin{list}{--}
      {
	\setlength{\leftmargin} {4mm}
	\setlength{\parsep}     {0pt}
	\setlength{\itemsep}    {0mm}
	\setlength{\topsep}     {\itemsep}
	\setlength{\partopsep}  {0pt}
	\setlength{\parskip}    {0pt}
	}} {
  \end{list}}
\begin{document}
\hspace*{13cm}{
\parbox[t]{5cm}{
YITP-SB-04-53\\
CERN-PH-TH/2004-193}}
\vskip 0.3cm
\title{Global Analysis of Neutrino Data\footnote{Talk given at
the Nobel Symposium on Neutrino Physics, 
Haga Slott, Enköping, Sweden, 
August 19-24, 2004}}
\author{M. C. Gonzalez-Garcia
\thanks{E-mail: concha@insti.physics.sunysb.edu}}
\affiliation{Theory Division, CERN, CH-1211, Geneva 23, Switzerland,\\
Y.I.T.P., SUNY at Stony Brook, Stony Brook, NY 11794-3840, USA,\\
IFIC, Universitat de Val\`encia - C.S.I.C., Apt 22085, 46071
Val\`encia, Spain}
\begin{abstract}
In this talk I review the present status of neutrino masses and mixing
and some of their implications for particle physics phenomenology.  
\end{abstract}
\maketitle
\section{Introduction: The New Minimal Standard Model}
The SM is based on the gauge symmetry
$SU(3)_{\rm C}\times SU(2)_{\rm L}\times U(1)_{\rm Y}$  
spontaneously broken to $SU(3)_{\rm C}\times U(1)_{\rm EM}$ by the 
the vacuum expectations value (VEV), $v$, of the a Higgs doublet 
field $\phi$.
The SM contains three fermion generations which reside in chiral 
representations of the gauge group. Right-handed fields are included 
for charged fermions as they are needed to build the electromagnetic
and strong currents.  No right-handed neutrino is included in the
model since neutrinos are neutral. 

In the SM, fermion masses arise from the Yukawa interactions which 
couple the right-handed fermion singlets to the left-handed fermion 
doublets and the Higgs doublet. After spontaneous electroweak symmetry 
breaking these interactions lead to charged fermion masses 
but leave the neutrinos massless. No Yukawa interaction can be written that
would give a tree level mass to the neutrino because no right-handed 
neutrino field exists in the model. 

One could think that neutrino masses could arise from loop corrections
if these corrections induced effective terms ${Y^\nu_{ij}\over v} 
\left(\bar L_{Li} \tilde \phi\right) 
\left( {\tilde\phi}^T L^C_{Lj}\right)$ where $L_{Li}$ are the lepton
doublets. 
This, however, cannot happen because  within the SM 
$G_{\rm SM}^{\rm global}=U(1)_B\times U(1)_e\times U(1)_\mu\times U(1)_\tau$ 
is an   accidental global symmetry.
Here $U(1)_B$ is the baryon number symmetry, and $U(1)_{e,\mu,\tau}$
are the three lepton flavor symmetries.
Terms of the form above violate $G_{\rm SM}^{\rm global}$ and 
therefore cannot be induced by loop corrections. 
Furthermore, they cannot be induced  by non-perturbative corrections 
because the $U(1)_{B-L}$ subgroup of $G_{\rm SM}^{\rm global}$ 
is non-anomalous.

It follows then that the SM predicts that neutrinos are {\sl strictly} 
massless. Consequently, there is neither mixing nor CP violation in the
leptonic sector. 

We now know that this picture cannot be correct. Over several years
we have accumulated important experimental evidence that neutrinos
are massive particles and there is mixing in the leptonic sector:
\begin{myitemize}
\item Solar $\nu_e's$ convert to $\nu_{\mu}$ or $\nu_\tau$ with 
confidence level (CL) of more than 7$\sigma$~\cite{sk,sno}.
\item KamLAND find that reactor $\overline{\nu}_e$ disappear over distances
of about 180 km and they observe a distortion of their energy spectrum.
Altogether their evidence has more than 3$\sigma$ CL~\cite{kamland}.
\item The evidence of atmospheric (ATM) $\nu_\mu$  disappearing is now at
$> 15 \sigma$, most likely converting to $\nu_\tau$~\cite{sk}.
\item  K2K observe the disappearance of accelerator $\nu_\mu$'s at 
distance of 250 km and find a distortion of their energy spectrum 
with a CL of 2.5--4 $\sigma$~\cite{sk,acc}.
\item  LSND found evidence for 
$\overline{\nu_\mu}\rightarrow\overline{\nu_e}$. This evidence has not
been confirm by any other experiment so far and it is being tested
by MiniBooNE~\cite{acc}. 
\end{myitemize}
These results imply that neutrinos are massive and the Standard Model
has to be extended at least to include neutrino masses. 
This minimal extension is what I call 
{\sl The New Minimal Standard Model}. 

In the New Minimal Standard Model flavour is mixed in the  
CC interactions of the leptons, and a leptonic mixing matrix
appears analogous to the CKM matrix for the quarks.
However the discussion of leptonic mixing is complicated
by two factors. First the number 
massive neutrinos ($n$) is unknown, since there are
no constraints on the number of right-handed, SM-singlet,
neutrinos. Second, since neutrinos carry neither 
color nor electromagnetic charge, they could be Majorana
fermions. As a consequence the number of new parameters
in the model depends on the number of massive neutrino states and on 
whether they are Dirac or Majorana particles.

In general, if we denote the neutrino mass eigenstates by $\nu_i$,
$i=1,2,\ldots,n$, and the charged lepton mass eigenstates 
by $l_i=(e,\mu,\tau)$, in the mass basis, leptonic CC interactions 
are given by
\begin{equation}
-{\cal L}_{\rm CC}={g\over\sqrt{2}}\, \overline{{l_i}_L} 
\, \gamma^\mu \, U_{ij}\,  \nu_j \; W_\mu^+ +{\rm h.c.}.
\label{CClepmas}  
\end{equation} 
Here $U$ is a $3\times n$ matrix 
$U_{ij}=P_{\ell,ii}\, {V^\ell_{ik}}^\dagger \, V^\nu_{kj}\, (P_{\nu,jj})$
where $V^\ell$ ($3\times 3$) and $V^\nu$ ($n\times n$) are the diagonalizing
matrix of the charged leptons and neutrino mass matrix respectively
${V^\ell}^\dagger M_\ell M_\ell^\dagger V^\ell=
{\rm diag}(m_e^2,m_\mu^2,m_\tau^2)$ and 
${V^\nu}^\dagger M_\nu^\dagger M_\nu V^\nu={\rm diag}(m_1^2,m_2^2,m_3^2,\dots,
m_n^2)$.
 
$P_\ell$ is a diagonal $3\times3$ phase matrix, that is conventionally
used to reduce by three the number of phases in $U$.
$P_\nu$ is a diagonal matrix with additional arbitrary phases (chosen
to reduce the number of phases in $U$) only for Dirac states. 
For Majorana neutrinos, this matrix is simply a unit matrix, 
the reason being that if one rotates a Majorana neutrino 
by a phase, this phase will appear in its
mass term which will no longer be real.
Thus, the number of phases that can be absorbed by redefining the
mass eigenstates depends on whether the neutrinos are Dirac or Majorana 
particles. In particular, if there are only three Majorana (Dirac) neutrinos, 
$U$ is a $3\times 3$ matrix analogous to the CKM matrix for the
quarks
but due to the Majorana (Dirac) nature of the neutrinos it depends on 
six (four) independent parameters: three mixing angles and three (one) phases.

A consequence of the presence of the leptonic mixing is the possibility 
of  flavour oscillations of the neutrinos. In this symposium we had 
a beautiful historical introduction to neutrino oscillations by S. 
Bilenky and two very interesting talks on the phenomenology of 
oscillations in matter
by A. Smirnov and  on three neutrino mixing effects by  E. Akhmedov.
So I will only briefly summarize here the elements which are relevant
for the phenomenological analysis that I will present. 

Neutrino oscillations appear because a neutrino of energy $E$ produced 
in a CC interaction with a charged lepton $l_\alpha$ can be detected 
via a CC interaction with a charged lepton $l_\beta$ with a probability
which presents an oscillatory behaviour, with oscillation lengths  
$L_{0,ij}^{\rm osc}=\frac{4 \pi E}{\Delta m_{ij}^2}$
and amplitude that is proportional to elements in the mixing matrix. 
Neutrino oscillations are only sensitive
to mass squared differences. Also, the Majorana phases cancel out and only the
Dirac phase is observable. 
Experimental information on absolute neutrino masses can be obtained
from Tritium $\beta$ decay experiments~\cite{mass} and from its effect
on the cosmic microwave background radiation and large structure formation
data~\cite{cosmo}. Also if neutrinos are Majorana particles their mass
and also additional phases can be determined in $\nu$-less $\beta\beta$
decay experiments~\cite{bb}. 

When neutrinos travel through regions of dense matter, they can undergo 
forward scattering with the particles in the medium. These interactions are, 
in general, flavour dependent and as a consequence 
the oscillation pattern described above is modified but it still
depends only on the mass squared differences  and it is 
independent of the Majorana phases.

The neutrino experiments described above have measured 
some non-vanishing $P_{\alpha\beta}$ and from these
measurements we have inferred all the positive evidence that 
we have on the non-vanishing values of neutrino masses and mixing. In 
the following I will derive the allowed ranges for the mass and mixing 
parameters when the bulk of data is consistently combined.
\section{Orthodox Fits}
I denote by {\sl Orthodox} fits those which try to explain the evidences
from solar, KamLAND, ATM and K2K experiments and assume that
the LSND evidence will not be confirmed by MiniBoone.
\subsection{Analysis of Solar and KamLAND}
\begin{figure}[t] 
\centerline{\psfig{figure=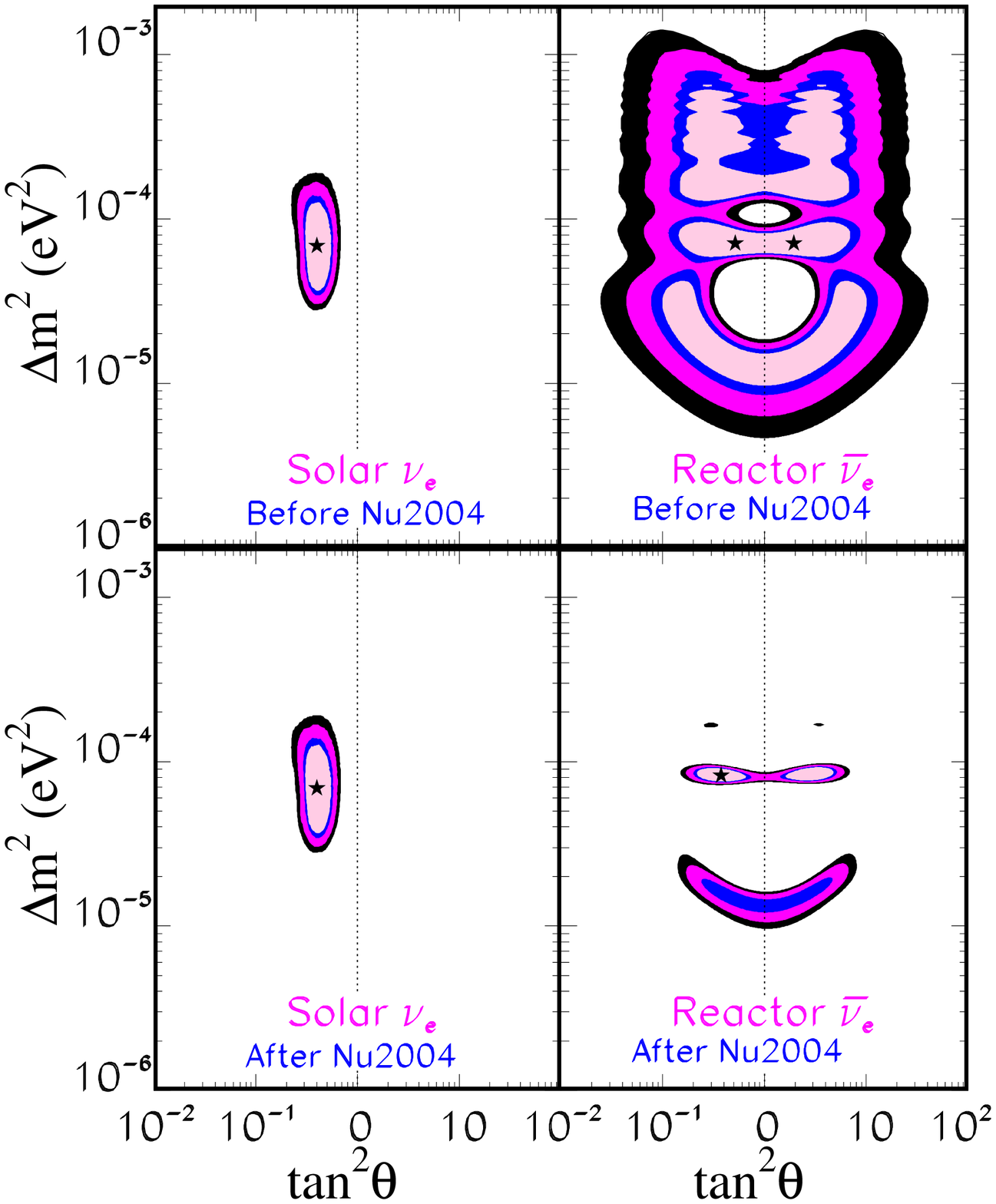,width=2.5in}}
\vspace*{-0.5cm}
\centerline{\psfig{figure=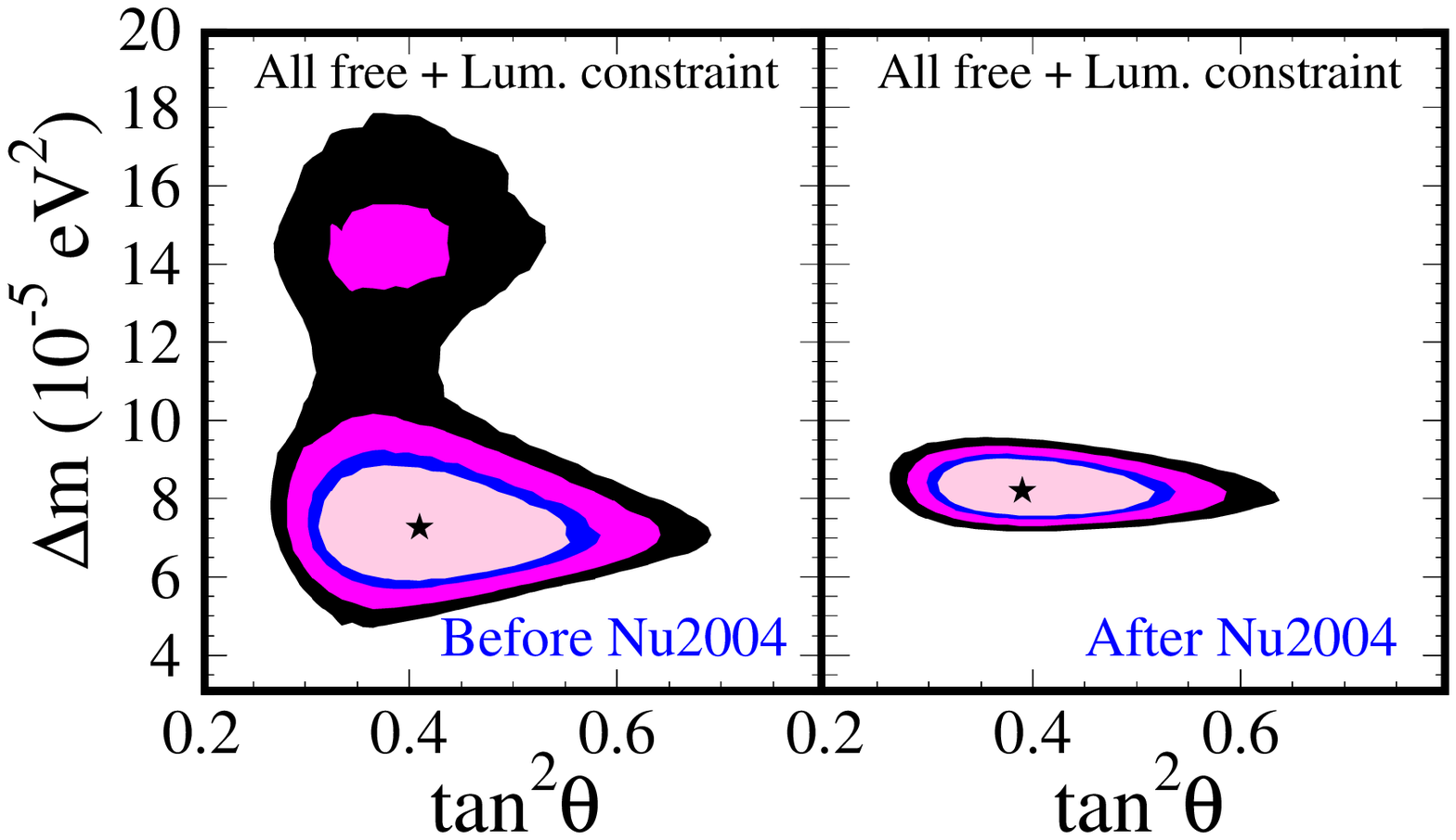,width=2.5in}}
\caption{
Allowed regions for 2-$\nu$ oscillations of 
solar $\nu_e$ and  KamLAND $\bar\nu_e$ (upper four panels), 
and for the combination of KamLAND and solar data under
the hypothesis of CPT conservation (lower two panels). 
The different contours correspond to the allowed regions 
at 90\%, 95\%, 99\% and $3\sigma$ CL.}
\label{fig:kland}
\end{figure}

In Fig.~\ref{fig:kland} I show the results from our latest analysis~\cite{oursolar} of KamLAND $\overline{\nu}_e$ disappearance data,
solar $\nu_e$ data and their combination under
the hypothesis of CPT symmetry. The main new ingredient 
is the inclusion of the new results from KamLAND presented in 
$\nu$-2004 conference in
June. We have also taken into account the new gallium measurement 
which leads to the average value $68.1\pm3.75$ SNU. 
The main features of these results are: 
\begin{myitemize}
  \item In the analysis of solar data, only LMA is allowed at more than
    $3\sigma$ and maximal mixing is rejected by the solar analysis at more
    than $5\sigma$. This is so since the release of the SNO salt-data
    (SNOII) in Sep 2003.   
  \item In the analysis of the new KamLAND data the 3$\sigma$ region does 
not extend to mass values larger than $\Delta m^2_{21}=2\times 10^{-4}$ 
eV$^2$ because for larger $\Delta m^2_{21}$ values, the  predicted
spectral distortions are too small to fit the new KamLAND data.
\item the combined analysis allows only the lowest LMA region 
at 3$\sigma$ with best-fit point and 1$\sigma$ ranges:
    \begin{equation}
	\Delta m^2=\left[8.2^{+0.3}_{-0.3}\right]
        \times 10^{-5}~ {\rm eV}^2 \,, 
	\tan^2\theta=0.39^{+0.05}_{-0.04}\,. 
    \end{equation}
\end{myitemize}
These results are in agreement with those reported in the several
state-of-the-art analysis of solar and KamLAND data which exist in the
literature.  

\subsection{Atmospheric and K2K Neutrinos}
\begin{figure}[t] 
\centerline{\psfig{figure=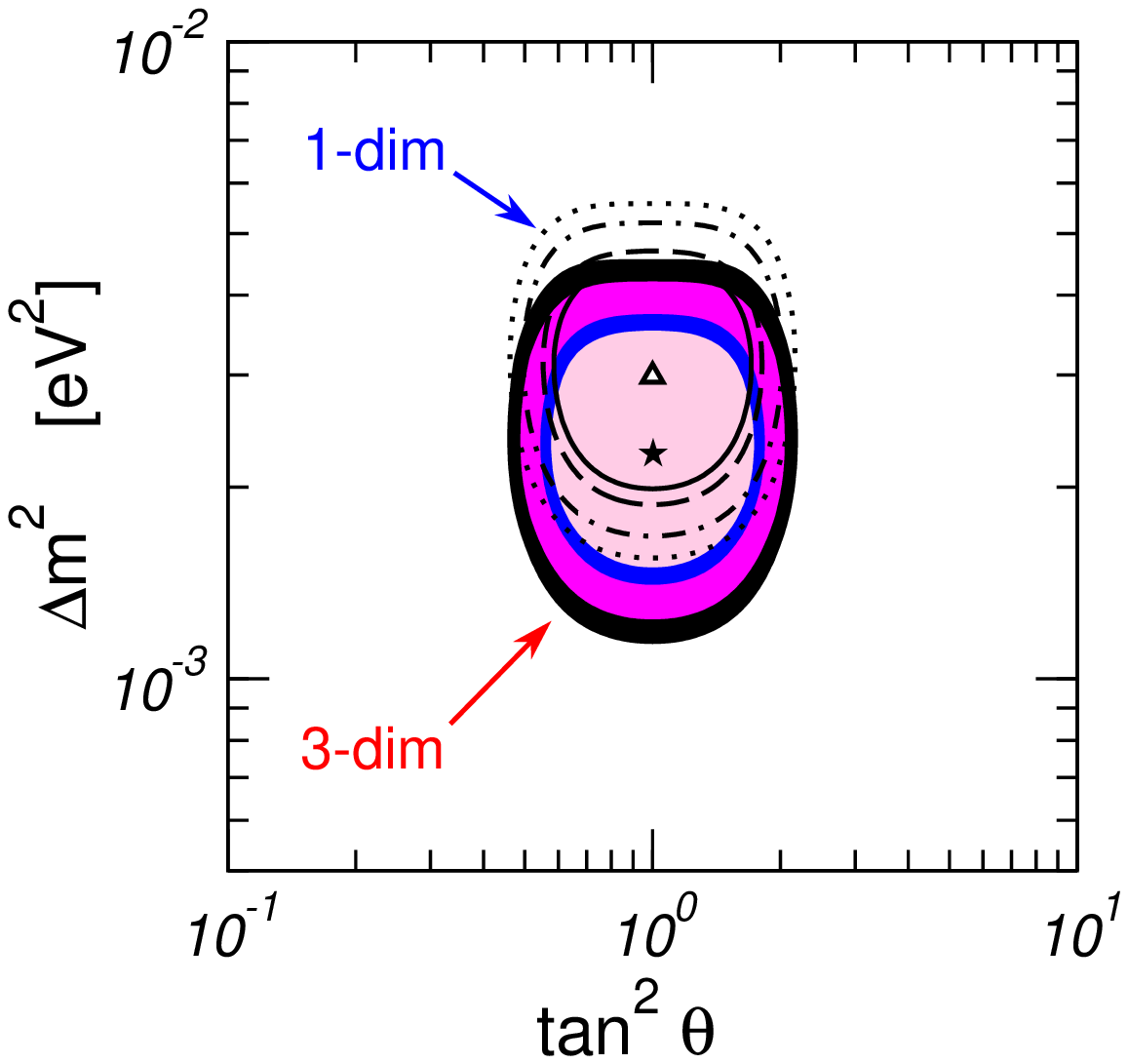,width=2.in}}
\vspace*{-0.5cm}
\centerline{\psfig{figure=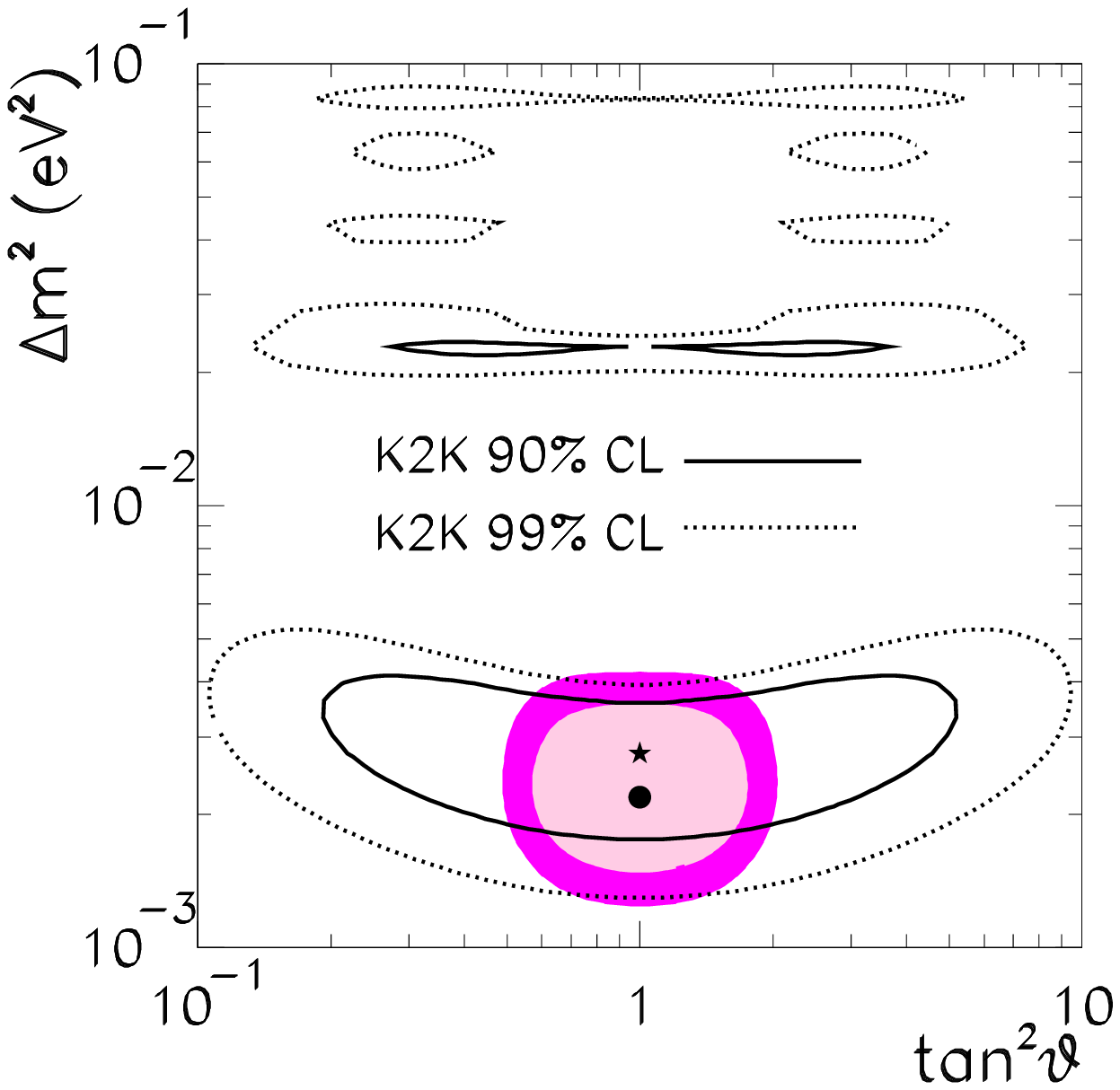,width=2.in}}
\caption{\textit{Upper}:
      Allowed regions from the analysis of
      ATM data using the new (full regions labeled ``3-dim'')
      and old (empty curves labeled as ``1-dim'') SK1 data and
      ATM fluxes. The different contours correspond to at
      90\%, 95\%, 99\% and $3\sigma$ CL.
\textit{Lower}:
      Allowed regions from the analysis of K2K data at 90\% (full line)
      and 99\% (dashed line) confidence level. For comparison we also show
      the corresponding allowed regions from ATM neutrinos 
      at the same CL.}
\label{fig:atm2fam} 
\end{figure}

In Fig.~\ref{fig:atm2fam} I show the results of our
latest analysis of the ATM neutrino data~\cite{ouratm}, 
which includes the full data set of Super-Kamiokande phase I (SK1)
as well as:
\begin{myitemize}
  \item use of new three-dimensional fluxes from Honda;
  \item improved interaction cross sections; 
  \item some improvements in the Monte-Carlo which lead to some
    changes in the actual values of the data points.
\end{myitemize}
Our results show good quantitative agreement with those of the SK
collaboration. In particular we find that after inclusion of the above
effects, the allowed region is shifted to lower $\Delta m^2$ with
best-fit point and 1$\sigma$ ranges:
\begin{equation}
    \Delta m^2 = \left[2.2^{+0.6}_{-0.4}\right]\times 10^{-3}~{\rm eV}^2\,, \qquad
    \tan^2\theta=1^{+0.35}_{-0.26} \,.
\end{equation}

At this point I would like to raise a word of caution.  In all present
analysis of ATM data, two main sources of theoretical flux
uncertainties are included: an energy independent normalization error
and a ``tilt'' error which parametrizes the uncertainty in the
$E^{-\gamma}$ dependence of the flux. Some additional uncertainties in
the ratios of the samples at different energies are also allowed as
well as uncertainties in the zenith dependences.  However, we still
lack a well established range of theoretical flux uncertainties within
a given ATM flux calculation, in a similar fashion to what it
is provided for the solar neutrino fluxes by the SSM. In the absence
of these, we cannot be sure that we are accounting for the most
general characterization of the energy dependence of the ATM
neutrino flux uncertainties.
Given the large amount of data points provided by the SK
experiment, this is becoming an important issue in the ATM
neutrino analysis. There is a chance that the ATM fluxes may
be still too ``rigid'', even when allowed to change within the
presently considered uncertainties. As a consequence, we may be
over-constraining the oscillation parameters. 

The evidence of oscillation of ATM $\nu_\mu$ has been now
confirmed by the long-baseline (LBL) K2K experiment which has observed 
not only a deficit of $\nu_\mu$'s at a distance of 250 km but
it has also measured the distortion of their energy spectrum.
In the lower panel of Fig.~\ref{fig:atm2fam} 
I show the results of our preliminary analysis
of the K2K data which graphically illustrates this agreement.
\subsection{Three-Neutrino Oscillations}
The minimum joint description of ATM, K2K, solar and reactor
data requires that all the three known neutrinos take part in the
oscillations.  The mixing parameters are encoded in the $3
\times 3$ lepton mixing matrix which can be conveniently parametrized
in the standard form $U=$
{\small
\begin{eqnarray}
\pmatrix{
    1&0&0 \cr
    0& {c_{23}} & {s_{23}} \cr
    0& -{s_{23}}& {c_{23}}\cr}
    \pmatrix{
    {c_{13}} & 0 & {s_{13}}e^{i {\delta}}\cr
    0&1&0\cr 
    -{ s_{13}}e^{-i {\delta}} & 0  & {c_{13}}\cr} 
\pmatrix{
    c_{21} & {s_{12}}&0\cr
    -{s_{12}}& {c_{12}}&0\cr
    0&0&1\cr}
    \nonumber
\end{eqnarray}}
where $c_{ij} \equiv \cos\theta_{ij}$ and $s_{ij} \equiv
\sin\theta_{ij}$. The angles $\theta_{ij}$ can be taken without loss of
generality to lie in the first quadrant, $\theta_{ij} \in [0,\pi/2]$.  

There are two possible mass orderings, which we denote as {\sl Normal}
and  {\sl Inverted}.  In the normal scheme $m_1<m_2<m_3$ while in 
the inverted one $m_3< m_1<m_2$. 

In total the 3-$\nu$ oscillation analysis involves seven
parameters: 2 mass differences, 3 mixing angles, the CP phase and the
mass ordering. Generic 3-$\nu$ oscillation effects
include:
\begin{myitemize}
  \item coupled oscillations with two different wavelengths;
  \item CP violating effects;
  \item difference between Normal and Inverted schemes.
\end{myitemize}
The strength of these effects is controlled by the values of the ratio
of mass differences $\alpha \equiv \Delta m^2_{21}/|\Delta m_{31}^2|$, by the
mixing angle $\theta_{13}$ and by the CP phase $\delta$. 

For solar and ATM oscillations,
\begin{equation} \label{eq:deltahier}
    \Delta m^2_\odot = \Delta m^2_{21} \ll 
|\Delta m_{31}^2|\simeq|\Delta m_{32}^2|
=\Delta m^2_{\rm atm}.
\end{equation}
and the joint 3-$\nu$ analysis simplifies as follows:
\begin{myitemize}
  \item for solar and KamLAND neutrinos, the oscillations with the
    ATM oscillation length are completely averaged and the
    survival probability takes the form:
    \begin{equation}
	P^{3\nu}_{ee}
	=\sin^4\theta_{13}+ \cos^4\theta_{13}P^{2\nu}_{ee} 
	\label{eq:p3}
    \end{equation}
    where in the Sun $P^{2\nu}_{ee}$ is obtained with the modified sun
    density $N_{e}\rightarrow \cos^2\theta_{13} N_e$. So the analyses
    of solar data constrain three of the seven parameters: $\Delta
    m^2_{21}, \theta_{12}$ and $\theta_{13}$. The effect of
    $\theta_{13}$ is to decrease the energy dependence of the 
    survival probability;
  \item for ATM and K2K neutrinos, the solar wavelength is too
    long and the corresponding oscillating phase is negligible. As a
    consequence, the ATM data analysis restricts $\Delta
    m^2_{31}\simeq \Delta m^2_{32}$, $\theta_{23}$ and $\theta_{13}$,
    the latter being the only parameter common to both solar and
    ATM neutrino oscillations and which may potentially allow
    for some mutual influence. The effect of $\theta_{13}$ is to add a
    $\nu_\mu\rightarrow\nu_e$ contribution to the ATM
    oscillations;
  \item at CHOOZ  the solar wavelength is unobservable 
    and the relevant survival
    probability oscillates with wavelength determined by $\Delta
    m^2_{31}$ and amplitude determined by $\theta_{13}$. 
\end{myitemize}

In this approximation, the CP phase is unobservable. In principle
there is a dependence on the Normal versus Inverted orderings due to
matter effects in the Earth for ATM neutrinos. However, this
effect is controlled by the mixing angle $\theta_{13}$. Presently 
all data favour small $\theta_{13}$ 
with best fit point very near $\theta_{13}=0$. The dominant
constraint arises from the combined analysis of CHOOZ reactor and
ATM data and it is further limited by the solar and 
KamLAND results. As a consequence, this effect is too small to be 
statistically meaningful in the present analysis. 
\begin{figure}[t] 
\centerline{\psfig{figure=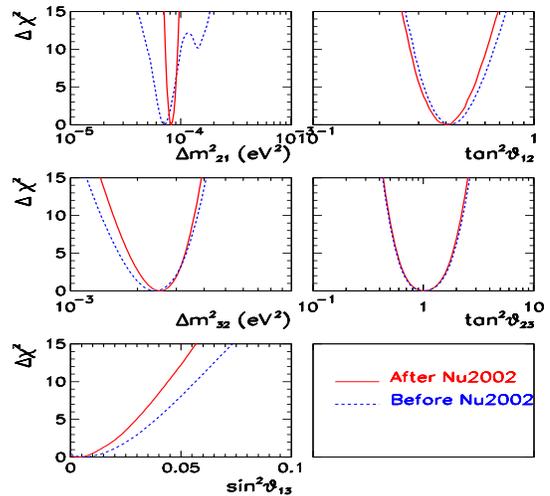,width=3in,height=2.7in}}
\caption{Global $3\nu$ oscillation analysis. Each panel on the left 
      shows the dependence of $\Delta\chi^2$ on each of the five
      parameters from the global analysis compared to the
      bound prior to the inclusion of the new KamLAND 
     and K2K data.}
\label{fig:chiglo}
\end{figure}

In Fig.~\ref{fig:chiglo} I show the individual bounds on each of the
five parameters derived from the global analysis.  To illustrate the
effect of the new KamLAND and K2K data we also show the results before their 
inclusion. In each panel the displayed $\Delta\chi^2$ has
been marginalized with respect to the undisplayed parameters.
As seen in the figure the main effect of the new data is a better determination
of the mass differences. A secondary effect is the slight improvement of
upper bound on $\theta_{13}$. This is mainly driven by K2K which 
favours the higher mass part of the ATM neutrino region for which
the corresponding bound from CHOOZ is tighter. 

Altogether we find the following $3\sigma$ ranges:
\begin{eqnarray} 
7.3 \leq {\displaystyle\frac{\Delta m^2_{21}}{{10^{-5}~{\rm eV}^2}}}\leq 9.3\,
&\qquad & 
    0.28 \leq  \tan^2\theta_{12}  \leq 0.60 \,, \nonumber    \\
1.6\leq {\displaystyle\frac{\Delta m^2_{32}}{{10^{-3}~{\rm eV}^2}}}\leq 3.6\,, 
&\qquad & 
    0.5 \leq \tan^2\theta_{23} \leq 2.1 \,, \label{eq:sranges} \\
&     &\sin^2\theta_{13}  \leq 0.041 \,. \nonumber
\end{eqnarray}
These results can be translated into our present knowledge of the
moduli of the mixing matrix $U$:
\begin{equation}
    |U| =\pmatrix{0.79-0.88&0.47-0.61&<0.20 \cr
0.19-0.52& 0.42-0.73&0.58-0.82\cr 
0.20-0.53&0.44-0.74&0.56-0.81\cr}\; .
\end{equation}

To finish this section I would like to discuss the possible observability 
of $\Delta m^2_{21}$ oscillations in ATM neutrinos. These effects
although small can, in principle be visible, mostly in the low energy
$\nu_e$ events provided that the mixing angle $\theta_{23}$ deviates
from maximal (see Ref.~\cite{nohier} and references in therein). 
As a matter of fact, they can lead to an increase or a decrease of the 
sub-GeV e-like events depending on the octant of the angle $\theta_{23}$
and they can be our best observable to detect both deviations of
$\theta_{23}$ from maximal as well as the sign of the deviation. 

\begin{figure}     
\centerline{\psfig{figure=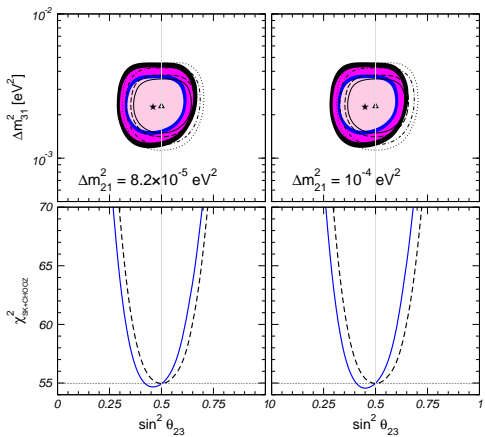,width=2.5in}}
    \caption{Effect of $\Delta m^2_{21}$ oscillations on 
      the allowed regions of the
      oscillation parameters $\Delta m^2_{31}$ and $\sin^2\theta_{23}$ from
      the combined analysis of all the ATM and CHOOZ data
      samples. In the lower panels we show the dependence of the
      $\chi^2$ function on $\theta_{23}$, marginalized with respect to
      $\Delta m^2_{31}$. Hollow and dashed black lines are for 
      $\Delta m^2_{21} =0$.}
\label{fig:present}
\end{figure}

The present data may already give some hint of deviation of the 2-3
mixing from maximal. Indeed, there is some excess of the $e-$like
events in the sub-GeV range. The excess increases with decrease of
energy within the sample as expected from a $\Delta m^2_{21}$ effect.
To illustrate this I show in Fig.~\ref{fig:present} the results of
the global analysis of ATM and CHOOZ data in the framework
of 3$\nu$ oscillations  taking into account also the effect of
$\Delta m^2_{21}$ oscillations~\cite{nohier}.

From the figure we see that, even with the present uncertainties, the
ATM data has some sensitivity to $\Delta m^2_{21}$ oscillation
effects and that these effects break the symmetry in $\theta_{23}$
around maximal mixing.
Although statistically not very significant, this
preference for non-maximal 2-3 mixing is a physical effect on the
present neutrino data, induced by the fact than an excess of events is
observed in sub-GeV electrons but not in sub-GeV muons nor, in the
same amount, in the multi-GeV electrons. 
As a consequence, this
excess cannot be fully explained by a combination of a global
rescaling and a ``tilt'', of the fluxes within the assumed 
uncertainties. 
\section{Unorthodox Fits}
I denote by {\sl Unorthodox} fits those in which either the matter contents
of the SM has been extended to include new light sterile neutrinos or
the basic symmetries of the model are violated, in most cases with the
aim of accommodating the LSND signal. 
\begin{figure}[t] 
\centerline{\psfig{figure=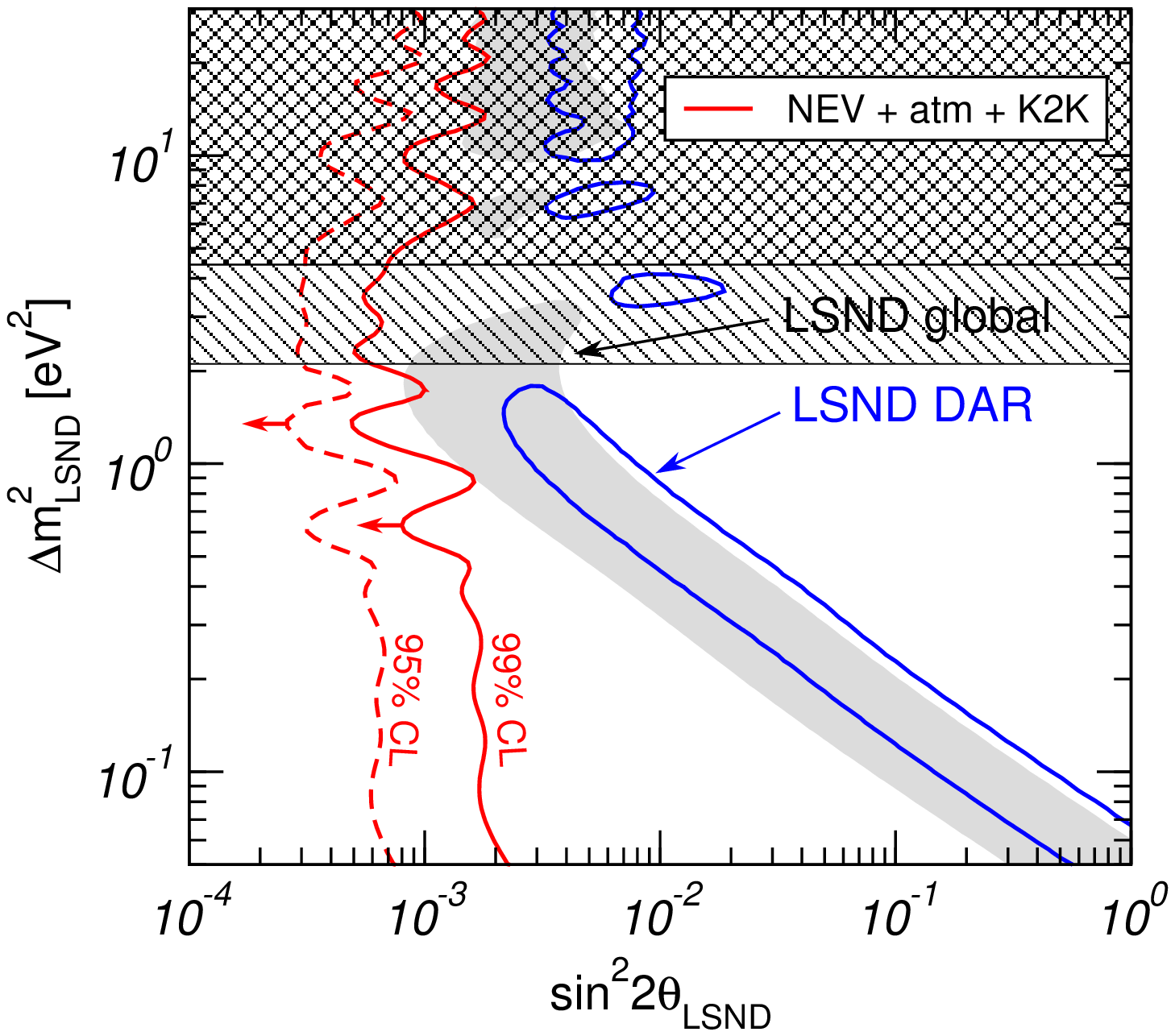,width=2.in}}
\vspace*{-0.2cm}
\centerline{\psfig{figure=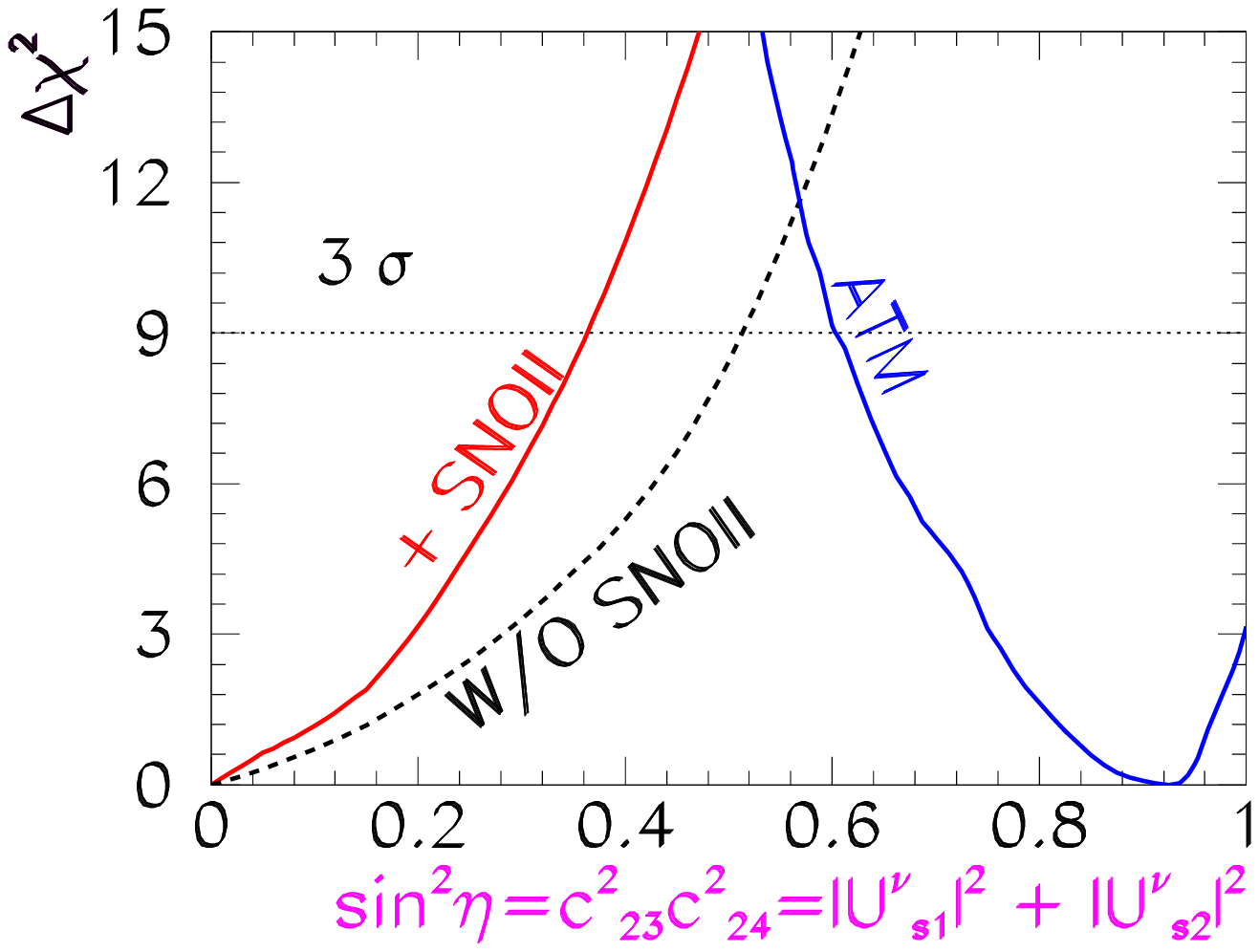,width=2.in}}
\caption{\textit{Upper}:
 Status of the 3+1 oscillation scenarios.
\textit{Lower}: Present status of the bounds on the
active-sterile admixture from solar and ATM neutrino
data in (2+2)-models.}
\label{fig:ster}
\end{figure}
\subsection{LSND and Sterile Neutrinos}
The LSND experiment 
found evidence of $\overline{\nu}_\mu \rightarrow \overline{\nu}_e$
neutrino conversion with $\Delta m^2\geq 0.1~{\rm eV}^2$. 
This result can be accommodated  together with those from solar,
reactor, ATM and LBL experiments 
into a single neutrino oscillation framework only if
there are at least three different scales of neutrino mass-squared
differences. 

As a first attempt to generate the required scales one can invoke 
the existence of a fourth light neutrino,
which must be {\it sterile} in order not to affect the invisible $Z^0$
decay width, precisely measured at LEP. 
There are six possible four-neutrino schemes which, in principle, 
can do the job. They can be divided in two classes: (3+1) and (2+2). In the
(3+1) schemes, there is a group of three close-by neutrino masses that
is separated from the fourth one by a gap of the order of 1~eV$^2$,
which is responsible for the SBL oscillations observed in the LSND
experiment. In (2+2) schemes, there are two pairs of close masses
separated by the LSND gap. The main difference between these two
classes is that in (2+2)-spectra 
the transition into the sterile neutrino is a solution of either the
solar or the ATM neutrino problem, or the sterile neutrino
takes part in both. This is not the case for a (3+1)-spectrum, where
the sterile neutrino could be only slightly mixed with the active ones
and mainly provide a description of the LSND result.

I show in Fig.~\ref{fig:ster} the latest results of the analysis of
neutrino data in these scenarios. 
The phenomenological situation at present is that none of the
four-neutrino scenarios are favored by the data. 
(3+1)-spectra are disfavored by the incompatibility between the
LSND signal and the negative results found by other short-baseline (SBL)
laboratory experiments. There is also a constraint on the possible
value of the heavier neutrino mass in this scenario from their
contribution to the energy density in the Universe which is presently
constrained by cosmic microwave background radiation and large scale
structure formation data~\cite{cosmo}. This is illustrated
in the upper panel of Fig.~\ref{fig:ster} taken 
from Ref.~\cite{michele} where they find 
that after the inclusion of the cosmological bound
there is only a marginal overlap at 99\% CL between the allowed LSND
region and the excluded region from SBL+ATM experiments. 
Concerning
(2+2)-spectra, they are ruled out by the existing constraints from
the sterile oscillations in solar and ATM data as illustrated
in the lower panel which shows that the
lower bound on the sterile component from the analysis of ATM
data and the upper bound from the analysis of solar data do not
overlap at more than $4\sigma$. 
\subsection{Neutrinos as Tests of Fundamental Symmetries}
\begin{figure}[t] \centering
\centerline{\psfig{figure=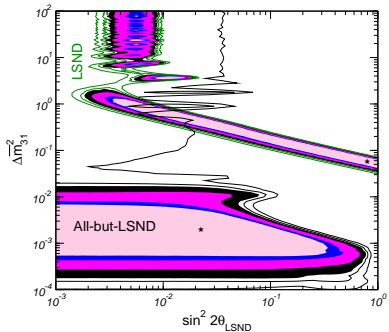,width=2.in}}
    \caption{
      90\%, 95\%, 99\%, and $3\sigma$ CL allowed regions (filled) in
      required to explain the LSND signal together with the
      corresponding allowed regions from our global analysis of
      all-but-LSND data. The contour lines correspond to $\Delta\chi^2
      = 13$ and 16 (3.2$\sigma$ and 3.6$\sigma$, respectively).}
\label{fig:cptstatus}
\end{figure}

Alternative explanations to the LSND result include the possibility of
CPT~\cite{CPT} violation, which implies that the masses and mixing
angles of neutrinos may be different from those of antineutrinos. 
To test this possibility, in Ref.~\cite{cpt} we 
performed an analysis of the existing data from solar,
ATM, LBL, reactor and SBL experiments in the
framework of CPT violating oscillations. The outcome of the analysis
is that, presently, the hypothesis of CPT violation is not supported
by the data. This arises from two main facts: (i) 
KamLand finds that reactor $\overline\nu_e$ oscillate with 
wavelength and amplitude in good agreement with the expectations 
from the LMA solution of the solar $\nu_e$; (ii) both ATM 
neutrinos and antineutrinos have to oscillate with similar wavelengths
and amplitudes to explain the ATM data. In general, as a result 
of these effects, the best fit to the data is very near CPT conservation 
and  in particular this rules out this scenario as explanation of the 
LSND anomaly. This is illustrated in Fig.~\ref{fig:cptstatus},
which shows clearly that there is no overlap below the $3\sigma$ level
between the LSND and the all-but-LSND allowed regions. 

Using the good description of neutrino data in terms of 
neutrino matter oscillations, it is also possible 
to constraint other exotic forms of new physics such as the violation 
of Lorentz Invariance (VLI)~\cite{VLI} 
induced by different asymptotic values of the velocity of the neutrinos, 
$c_1\neq c_2$, or  the violation of the 
equivalence principle  (VEP) ~\cite{VEP} due to non
universal coupling of the neutrinos, $\gamma_1\neq \gamma_2$ 
to the local gravitational potential,  among others.
These forms of new physics, if non-universal, can also induce 
neutrino flavour oscillations whose main differentiating characteristic
is a different energy dependence of the oscillation wavelength. 
For example for both VLI and VEP the oscillation wavelength decreases
with energy unlike for mass oscillations. ATM neutrino events
extend over several decades in energy. As a consequence they can test 
the presence of this effect even at the subdominant level. 
In Ref.~\cite{ouratm} we have performed an analysis of ATM 
and LBL neutrino data in terms of neutrino mass oscillations 
plus these new physics effects and we have concluded that the determination 
of mass and mixing parameters is robust under the presence of these unknown 
forms of new physics. Conversely, the analysis permits to impose strong 
constraints on the violations of these symmetries. For instance we find that
at 90\% CL the possible VLI a and VEP are limited to  
\begin{eqnarray} 
{\displaystyle\frac{|\Delta c|}{c} }
\leq 8.1\times 10^{-25} \;, &\qquad & 
    |\phi\, \Delta \gamma| \leq 4.0\times 10^{-25}\;. 
\end{eqnarray}
which constitute the strongest constraints on the violation of these 
fundamental symmetries.  
 
This work was supported in part by the National Science Foundation
grant PHY-0354776. M.C.G.-G.\ is also supported by Spanish Grants No
FPA-2001-3031 and GRUPOS03/013.

\end{document}